\documentclass[prb,showpacs,twocolumn,amsmath,amssymb]{revtex4}
\usepackage{graphicx}
\usepackage{dcolumn}
\usepackage{bm}
\begin{document}
\par
\title{Electron-beam-induced shift in the apparent position of a 
pinned vortex  in a thin superconducting film }
\author{John R. Clem}
\affiliation{Ames Laboratory - DOE and Department of Physics and
Astronomy, Iowa State University, Ames Iowa 50011 }
\date{\today}
\begin{abstract}
When an electron beam strikes a superconducting thin film near a pinned
vortex, it locally increases the temperature-dependent London penetration
depth and perturbs the circulating supercurrent, thereby distorting the
vortex's magnetic field toward the heated spot.  This phenomenon has been
used to visualize vortices pinned in SQUIDs using low-temperature scanning
electron microscopy. In this paper I develop a quantitative theory to
calculate the displacement of the vortex-generated magnetic-flux
distribution as a function of the distance of the beam spot from the
vortex core.  The results are calculated using four
different models for the spatial distribution of the thermal
power deposited by the electron beam.  
\end{abstract}
\pacs{\bf 74.78.-w, 74.25.Qt, 85.25.Dq}
\maketitle
\section{Introduction}
An important fundamental property of superfluids and
superconductors is that they admit quantized vortices, produced either by
rotation of neutral superfluids or by applying magnetic fields to
superconductors. Numerous experimental tools have been used to
visualize these vortices.  In neutral superfluids, charge
decoration has been used to observe vortices in superfluid
helium,\cite{Packard82} and ballistic expansion to observe them in
Bose-Einstein condensates.\cite{Zwierlein05} In superconducting films,
while scanning tunneling microscopy\cite{Hess89} detects quasiparticles in
the vortex core, the localized magnetic-field distributions generated by
the circulating supercurrents open the possibility of additional
techniques for the observation of singly quantized vortices, such as
Bitter decoration,\cite{Essmann67} magnetic force microscopy,
\cite{Moser95,Yuan96} scanning SQUID microscopy,\cite{Kirtley95} scanning
Hall-probe microscopy,\cite{Chang92,Oral96a,Oral96b}
magneto-optical detection,\cite{Goa03} Lorentz
microscopy,\cite{Harada92} and electron holography.\cite{Bonevich93} 
Low-temperature scanning electron microscopy
(LTSEM)\cite{Clem80,Huebener84,Bosch85,Mannhart87,Gross94,Doderer97} and
laser scanning microscopy\cite{Scheuermann83,Lhota83} have been used to
visualize vortices in Josephson junctions.

Recently LTSEM has been used in a new
way\cite{Straub01,Doenitz04,Doenitz06} to detect the presence of pinned
vortices in thin-film superconducting quantum interference devices
(SQUIDs) at 77 K.  When the scanning electron beam strikes the film near 
a pinned vortex, it locally raises the temperature, decreases the
superfluid density, and increases the
temperature-dependent London penetration depth
$\lambda(T)$. The supercurrent circulating around the vortex is perturbed,
the perturbation being a dipole-like backflow current distribution, which
generates a corresponding magnetic-field perturbation.  As a result, the
overall magnetic-field distribution generated by the vortex is no longer
centered on the pinned vortex core but is distorted in the direction of
the beam spot.  The shift in the apparent position of the vortex
produces a small change in the return magnetic flux threading the SQUID's
central hole.  As the electron beam
rasters across the sample, the SQUID output, which is extremely
sensitive to such flux changes, can be displayed on a video screen.  The
resulting image reveals each pinned vortex as a pair of bright and dark
spots centered on the vortex, the bright (dark) spot corresponding to an
increase (decrease) in the vortex-generated magnetic flux sensed by the
SQUID. Since a quantitative description of
the above behavior is still lacking,  I develop in Sec.~II a theory to
calculate the displacement of the vortex-generated magnetic-flux
distribution as a function of the distance of the beam spot from the
vortex core. I present results using four models for the spatial
distribution of the thermal power deposited by the electron beam.  In
Sec.~III, I summarize the theoretical results and discuss possible
experiments to test the theory.

\section{Theory}

Consider a vortex pinned at the origin in an infinite
film of thickness $d$ less than the temperature-dependent London
penetration  depth $\lambda(T)$, such that
the relevant screening length is the Pearl length\cite{Pearl64}
$\Lambda(T) = \lambda^2/d$. Since the current density $\bm j$ is nearly
constant across the thickness, we need only consider the sheet-current
density
$\bm K = \bm j d$.  The magnetic induction
generated by the vortex  $\bm B(\bm \rho,z) = \nabla \times \bm A$ is
described by a vector potential $\bm A(\bm \rho,z)$ that in the plane of
the film ($z = 0$) obeys the London equation\cite{London61}
\begin{equation}
\bm A(\bm \rho,0) + \mu_0 \Lambda \bm K(\bm \rho) =
-\frac{\phi_0}{2\pi}\nabla \gamma = \hat \phi \frac{\phi_0}{2\pi \rho},
\label{London}
\end{equation}
where $\bm K(\bm \rho) = (2/\mu_0)\hat z \times \bm B(\bm \rho,0^+)$,
$\phi_0 = h/2e$ is the superconducting flux quantum,
$\gamma$ is the phase of the superconducting order parameter, $\bm \rho =
x \hat x +
y \hat y
$,  
$\rho =
\sqrt {x^2+y^2}$, $\hat \rho = \bm \rho/\rho$, and $\hat \phi = \hat z
\times \hat \rho$.  When the temperature of the film is spatially uniform
($T = T_0$), so is the Pearl length $\Lambda = \Lambda(T_0) =
\Lambda_0$, and the solution of Eq.\ (\ref{London}) is well
known.  The sheet-current density is \cite{Pearl64,deGennes66}
\begin{equation}
\bm K_0(\bm \rho)=\frac{-i \phi_0}{2 \pi^2 \mu_0}\int \frac{\hat
q_\perp}{1+2q\Lambda_0} e^{i\bm q \cdot \bm\rho} d^2 q,
\label{K0}
\end{equation}
where $\bm q = q_x \hat x  +  q_y \hat y$, $q = \sqrt{q_x^2 + q_y^2},$ 
$\hat q =
\bm q/q$, 
and $\hat q_\perp = \hat z \times \hat q.$ For $\rho \ll \Lambda_0$, 
\begin{equation}
\bm K_0(\bm \rho) 
\approx \hat \phi \phi_0/2 \pi
\mu_0 \Lambda_0 \rho, 
\label{K0<}
\end{equation}
while for  $\rho \gg \Lambda_0$,
\begin{equation}
\bm K_0(\bm \rho) \approx \hat \phi \phi_0/\pi
\mu_0 \rho^2.
\label{K0>}
\end{equation}
The corresponding magnetic
field distribution $\bm B_0(\bm \rho,z)$  is centered on the $z$ axis;
thus
$\int B_{0z}(\bm \rho,z)d ^2 \rho = \phi_0$ and $\int \bm \rho
B_{0z}(\bm \rho,z)d ^2 \rho = 0$. Above the film at a distance
$r =
\sqrt{\rho^2 + z^2}$ somewhat greater than
$\Lambda_0$, $\bm B_0=\nabla \times \bm A_0$ appears as if produced by a
magnetic monopole at the origin; i.e., $\bm B_0 \approx
\hat r \phi_0/2\pi r^2$, where $\bm r = \bm \rho +  z \hat z$ and $\hat r
=
\bm r/r$. 

When a sharply focused  electron beam  scans across the
superconducting  film, depositing  thermal power $P_0$ in the
film and its substrate, this locally raises the film's temperature around
the beam spot at
$\bm
\rho_0(t) = x_0(t) \hat x + y_0(t) \hat y$. As in Refs.\
\onlinecite{Straub01,Doenitz04,Doenitz06}, I consider slow scans such
that  the temperature increment quasistatically follows the electron beam
and can be written as $\delta T(\bm \rho - \bm \rho_0)$.
This temperature increment can be calculated by
solving the steady-state heat diffusion equation, but this generally requires
a detailed knowledge of the initial pear-shaped spatial distribution of the
intensity of thermal energy deposition by the e-beam (i.e., the size and
shape of the interaction volume), the thermal conductivities $\kappa_F$
and $\kappa_S$ of the superconducting film and substrate, and the
coefficient of heat transfer from the film into the substrate
$\alpha_s$.\cite{Clem80,Gross94}  In the experiments of Ref.\
\onlinecite{Doenitz06} done at 77 K, the authors estimated that the
maximum temperature increment
$\Delta T = \delta T(0)$ was only few K. I thus consider
only small values of $\Delta T$, such that the deviation of
$\Lambda$ from $\Lambda_0$, i.e., $\delta \Lambda =
\Lambda_1(\bm
\rho) = (d\Lambda/dT)\delta T(\bm \rho)$, 
is a small perturbation. What is formally needed in the following
theory is
\begin{equation}
\delta {\tilde T}(\bm q) = \int \delta T(\bm \rho) e^{-i\bm q \cdot \bm
\rho} d^2
\rho,
\label{dTq}
\end{equation}
the two-dimensional Fourier
transform of the temperature increment.\cite{footnote}

To solve for the shift in the apparent position of the pinned vortex
when $\Delta T \ll T_0$, it is appropriate to use first-order
perturbation theory, with $\Lambda = \Lambda_0 + \Lambda_1$ and $\bm
A(\bm \rho,z) = A_0(\bm \rho,z) + A_1(\bm \rho,z),$ where quantities with
the subscript 1 are proportional to $\Delta T$.  Equation (\ref{London})
yields the boundary condition
\begin{equation}
\bm A_1(\bm \rho,0) + \mu_0 \Lambda_0 \bm K_1(\bm \rho) =
-\mu_0 \Lambda_1(\bm \rho - \bm \rho_0) \bm
K_0(\bm \rho).
\label{London1}
\end{equation}
It is useful to introduce the Green
functions $
\bm a_n(\bm r)$, $\bm b_n(\bm r) = \nabla \times \bm a_n(\bm r)$, and
$\bm k_n(\bm \rho) = (2/\mu_0) \hat z \times \bm b_n(\bm \rho, 0^+)$,
which obey
\begin{equation}
\bm a_n(\bm \rho,0) + \mu_0 \Lambda_0 \bm k_n(\bm \rho) = 
-\delta(\bm \rho) \hat x_n,
\end{equation}
where $\hat x_1 = \hat x,\; \hat
x_2 = \hat y,$ $\bm r = (\bm \rho,z)$, and $n =$ 1 or
2.  The solutions are
\begin{eqnarray}
\bm a_n(\bm \rho,z)\!\!\! &=&\!\!\!-\!\!\!\int \!\!\frac{d^2q}{(2\pi)^2}
\{[\hat q + i {\rm s}(z) \hat z](\hat q \cdot \hat x_n) \nonumber \\
&&+\frac{\hat q_\perp (\hat q_\perp\!\! \cdot \!\hat
x_n)}{1+2q\Lambda_0}\} e^{i\bm q \cdot \bm \rho} e^{-q|z|},\\
\bm b_n(\bm \rho,z)\!\!\! &=&\!\!\!-\!\!\!\int\!\! \frac{d^2q}{(2\pi)^2}
\frac{q[\hat q 
{\rm s}(z)\!+\!i\hat z] (\hat q_\perp\!\! \cdot\! \hat
x_n)}{1+2q\Lambda_0} e^{i\bm q
\cdot \bm \rho} e^{-q|z|},\\
\bm k_n(\bm \rho)\!\!\! &=&\!\!\!- \frac{2}{\mu_0}\!\int
\!\!\frac{d^2q}{(2\pi)^2}
\frac{q\hat q_\perp (\hat q_\perp\! \!\cdot\! \hat x_n)}{1+2q\Lambda_0}
e^{i\bm q \cdot \bm \rho},
\end{eqnarray}
where s($z$) = +1 if $z > 0$, 0 if $z=0$, and -1 if $z <0$.
Then Eq.\ (\ref{London1}) is solved by 
\begin{eqnarray}
\bm A_1(\bm \rho,z) &=& \sum_{n=1}^2 \int  D_n(\bm \rho',\bm \rho_0)
\bm a_n(\bm \rho - \bm \rho',z) d^2 \rho',\\
\bm B_1(\bm \rho,z) &=& \sum_{n=1}^2 \int  D_n(\bm \rho',\bm \rho_0)
\bm b_n(\bm \rho - \bm \rho',z) d^2 \rho',
\label{B1}\\
\bm K_1(\bm \rho) &=& \sum_{n=1}^2 \int  D_n(\bm \rho',\bm \rho_0)
\bm k_n(\bm \rho - \bm \rho') d^2 \rho',
\end{eqnarray}
where
\begin{equation}
\bm D(\bm \rho,\bm \rho_0) 
= \sum_{n=1}^2 D_n(\bm \rho,\bm \rho_0) \hat x_n
=\mu_0
\Lambda_1(\bm \rho - \bm
\rho_0)
\bm K_0(\bm \rho).
\label{D}
\end{equation}
Note that $\nabla \cdot \bm k_n(\bm \rho,z) = 0$ and 
$\nabla \cdot \bm K_1(\bm \rho,z) = 0$ but that 
$\nabla \cdot \bm a_n(\bm \rho,z) \ne 0$ and 
$\nabla \cdot \bm A_1(\bm \rho,z) \ne 0$.

Two important properties of the $z$ component of $\bm b_n(\bm \rho,z)$
are that $\int b_{nz}(\bm \rho,z) d^2 \rho = 0$ and 
\begin{equation}
\int \bm
\rho b_{nz}(\bm
\rho,z) d^2 \rho = \sum_{m=1}^2 \delta_{nm} \hat x_m \times \hat z.
\end{equation} 
Equations (\ref{B1}) and (\ref{D}) 
yield 
$\int B_{1z}(\bm \rho,z) d^2 \rho = 0$ and 
\begin{equation}
\int \bm
\rho B_{1z}(\bm
\rho,z) d^2 \rho = \int \bm D(\bm \rho,\bm \rho_0) d^2
\rho \times \hat z .
\end{equation}

Using the above properties of $\bm B_0$ and $\bm B_1$, we find that to 
first order in
$\Delta T$ the
electron-beam-induced shift in the apparent position of the vortex [i.e.,
the center of the distribution of $B_z(\bm \rho,z)$] is
\begin{equation}
\bm S(\bm \rho_0) = \frac{\int \bm \rho B_{z}(\bm \rho,z)
d^2 \rho} {\int B_{z}(\bm \rho,z) d^2 \rho}
=\frac{1}{\phi_0}\int \bm D(\bm \rho,\bm \rho_0) d^2
\rho \times \hat z .
\label{S}
\end{equation}
Using Eqs.\ (\ref{K0}), (\ref{dTq}), and (\ref{D}) and the properties of
two-dimensional Fourier transforms, we obtain the
following general expression for the shift\cite{footnote}
\begin{equation}
\bm S(\bm \rho_0) = \frac{\hat \rho_0}{\pi} \frac{d\Lambda}{dT}
\int_0^\infty \frac{\delta \tilde T(q)}{1+2q \Lambda_0}J_1(q\rho_0)qdq
\label{Sf}
\end{equation}
as a function of the beam spot position $\bm \rho_0 = x_0 \hat x + y_0
\hat y$,  where $\rho_0 = \sqrt{x_0^2 + y_0^2}$, $\hat \rho_0 = 
\bm \rho_0/\rho_0$, and $J_n$ here and in later equations is a Bessel
function of the first kind of order $n$.  Before we can evaluate the
integral in Eq.~(\ref{Sf}), we need to obtain an expression for 
$\delta
\tilde T(q)$ that provides a good description of the
experimental conditions.

Let us consider an experimental situation close to that described in
Ref.~\onlinecite{Doenitz06}, in which a well-collimated electron beam of
energy
$E_0 = 10$ keV, beam current $I_b$ = 7 nA, and radius $a \approx$ 5 nm was
normally incident upon an epitaxial c-axis-oriented superconducting
YBa$_2$Cu$_3$O$_7$ (YBCO) thin film of thickness $d = 80$ nm on a thick
SrTiO$_3$ substrate at 77 K.  The range
$R$, i.e., the maximum distance from the point of incidence the electron
travels until it thermalizes, was estimated to be $R
\approx$ 0.53 $\mu$m.  Each incident electron undergoes numerous
elastic and inelastic scattering processes as it loses energy along its
diffusion path $s$, and some of this energy emerges from the top surface of
the sample as back-scattered electrons, secondary electrons, x-rays, and
electromagnetic radiation.\cite{Reimer98}  The thermal power delivered to
the sample is $P_0 = fI_bE_0/e$, where the fraction of energy that is
ultimately converted into heat is of the order of $f$ =
40-80\%.\cite{Reimer98} 

Experiments and Monte Carlo
simulations  investigating the electron-scattering and
energy-loss processes show that the electrons follow relatively straight
paths when they are incident upon low-$Z$ targets such as carbon or
plastic, such that the diffusion cloud resembles a paint
brush.\cite{Reimer93,Reimer98}  On the other hand, when the electrons are
incident upon high-$Z$ targets such as gold, the more frequent large-angle
elastic scattering processes cause the diffusion cloud to be
apple-shaped.\cite{Reimer93, Reimer98}  The highest density of thermal
energy deposition is at the center of the beam spot, where all
the diffusion paths begin.    Because of the complexity of all the electron
scattering and diffusion processes, it is not possible to describe the
density of thermal power deposition
$P_v(\rho,z)$ with high accuracy by a simple analytic
formula.\cite{Reimer98}  However, because a mathematical expression for
$P_v(\rho,z)$  is needed for a calculation of
$\delta
\tilde T(q)$, which appears in the expression for
$\bm S(\bm \rho_0)$ in Eq.~(\ref{Sf}), I shall present calculations for
four different approximate models.  Each model assumes the
electron beam is centered on the
$z$ axis,
$P_v(\rho,z)$ is cylindrically  symmetric about the $z$ axis, and the volume
integral of 
$P_v(\rho,z)$  is equal to
$P_0$.

First, however, it is useful to consider how to solve for  $\delta
\tilde T(q)$ in general for any given density of thermal power deposition
$P_v(\rho,z)$.  At temperatures in the vicinity of 77 K and above, the
temperature perturbation
$F(\rho,z)$ in the steady state can be obtained by solving the thermal
diffusion equation\cite{Gross94}
$-\kappa
\nabla^2 F = P_v$ subject to the boundary condition that $\partial
F/\partial z = 0$ at the surface $z = 0$.  This assumes that to good
approximation the thermal conductivities of the superconductor and the
substrate are the same and equal to $\kappa$. The general solution is
\begin{equation}
F(\rho,z)= \frac{1}{2\pi}\int_0^\infty\tilde F(q,z)J_0(q\rho)q
dq,
\label{Frhoz}
\end{equation}
where
\begin{eqnarray}
\tilde F(q,z) = \frac{\pi}{\kappa q}\int_0^\infty d\rho'
\rho'\int_{-\infty}^0
dz'P_v(\rho',z')J_0(q\rho') \nonumber \\
\times (e^{-q|z-z'|}+e^{-q|z+z'|}).
\end{eqnarray}
The temperature perturbation at the surface is $\delta T(\rho) = F(\rho,0)$
or
\begin{equation}
\delta T(\rho)= \frac{1}{2\pi}\int_0^\infty \delta \tilde
T(q)J_0(q\rho)q
dq,
\end{equation}
where
\begin{equation}
\delta \tilde T(q)\! =\! \frac{2\pi}{\kappa q}\int_0^\infty\!\!\! d\rho'
\rho'\!\!\int_{-\infty}^0\!\!\!
dz'P_v(\rho',z')J_0(q\rho') 
e^{-q|z'|}.
\end{equation}
In the limit as $q \rightarrow 0,$ $\delta \tilde T(q)$ obeys
\begin{equation}
\delta \tilde T(q) \rightarrow \frac{2\pi}{\kappa q}\int_0^\infty d\rho'
\rho'\int_{-\infty}^0
dz'P_v(\rho',z')=\frac{P_0}{\kappa q},
\end{equation}
and in the limit as $\rho \rightarrow \infty,$ 
\begin{equation}
\delta T(\rho) \rightarrow \frac{P_0}{2\pi \kappa \rho},
\end{equation}
where $P_0$ is the total thermal power absorbed by the sample.
For each of the models discussed below, the common parameter is $R$, the
maximum range of the electron, and it is convenient for later use to
express $\delta \tilde T(q)$ and $\delta T(\rho)$ in terms  of the
dimensionless auxiliary functions $h(u)$ and $g(\tilde \rho)$ via
\begin{equation}
\delta \tilde T(q) = \frac{P_0}{\kappa q} h(u),
\label{dTh}
\end{equation}
where $u = qR$, and 
\begin{equation}
\delta T(\rho) = \frac{P_0}{2 \pi \kappa R} g(\tilde \rho),
\label{dTg}
\end{equation}
where $\tilde \rho = \rho/R$,
\begin{eqnarray}
h(u)&=&u\int_0^\infty J_0(u\tilde \rho)g(\tilde \rho)\tilde \rho d\tilde
\rho,\\ h(0) &=& 1, \\
g(\tilde \rho)&=&\int_0^\infty J_0(\tilde \rho u)h(u)du,\\
g(0)&=&\int_0^\infty h(u) du,
\end{eqnarray}
such that 
\begin{equation}
\delta T(0) = \frac{P_0}{2 \pi \kappa R} g(0).
\label{dT0}
\end{equation}

{\bf Model KO}.
To approximate sample heating effects in low-$Z$ materials such as
carbon, in which the penetrating electron
beam remains relatively straight,  Kanaya and Ono\cite{Kanaya84}
introduced a model in which thermal power is deposited with uniform
density throughout a cylinder of radius equal to the the beam radius
$a$ and length equal to the electron range $R$,
\begin{eqnarray}
P_{v,KO}(\rho,z)& =& P_0/\pi a^2 R, \rho < a, -R< z<0,\\
&=&0, {\rm otherwise}.
\end{eqnarray}
The function $h(u)$ defined in Eq.~(\ref{dTh}) becomes 
\begin{equation}
h_{KO}(u) = \frac{J_1(\alpha u)}{(\alpha u/2)} \frac{(1-e^{-u})}{u},
\label{hKO}
\end{equation}
where $\alpha = a/R$,
and the steady-state temperature increase at the surface is given by
Eq.~(\ref{dTg}), where 
\begin{equation}
g_{KO}(\tilde \rho) = \frac{4}{\pi
\alpha^2}\int_0^1 d\tilde z' 
\int_0^\alpha d\tilde \rho'\frac{\tilde \rho' \bm K(k)}
{\sqrt{(\tilde \rho + \tilde \rho')^2+\tilde z'^2}},
\end{equation}
and $\bm K(k)$ is the complete elliptic integral of the first kind of
modulus
\begin{equation}
k=\frac{2\sqrt{\tilde \rho \tilde \rho'}}{\sqrt{(\tilde \rho + \tilde
\rho')^2+\tilde z'^2}}.
\end{equation}
At the center of the beam spot,
\begin{equation}
g_{KO}(0) =
\ln\Big(\frac{1+\sqrt{1+\alpha^2}}{\alpha}\Big)
+\frac{(\sqrt{1+\alpha^2}-1)}{\alpha^2}.
\end{equation}

{\bf Model C}.
To approximate sample heating effects in higher-$Z$ materials in which the
diffusing electron beam spreads out radially from the center of the beam
spot into
$2\pi$ steradians, one may assume that the thermal power is deposited
with the following density,
\begin{eqnarray}
P_{v,C}(r)& =& P_0/2\pi a^2 R', r \le a,\\
&=& P_0/2\pi r^2 R' ,a \le r \le R,\\
&=&0, r\ge R,
\end{eqnarray}
where $R' = R-2a/3$.  This corresponds to the assumption that the maximum
density of deposited thermal power occurs at the center of the beam spot
but that the diffusing electrons lose energy at a constant rate $-dE/dr$ as
they move radially outward, depositing equal amounts of thermal energy into
hemispherical shells of increasing volume $2\pi r^2 dr$ until the electrons
thermalize at $r = R$.
The function $h(u)$ defined in Eq.~(\ref{dTh}) becomes 
\begin{eqnarray}
h_{C}(u)\!\!&=\!\!&
1-\frac{1}{(1\!\!-\!\!2\alpha/3)}\{u[J_0(u)-(2\alpha^2/3)J_0(\alpha
u)]
\nonumber \\
&&\!\!-[J_1(u)-(2\alpha/3) J_1(\alpha u)]-J_2(\alpha u)/3u
\nonumber \\
&&\!\!-(\pi u/2)[J_0(u)\bm H_1(u)-J_1(u)\bm H_0(u)] \nonumber \\
&&\!\!+(\pi u \alpha^2/3)[J_0(\alpha u)\bm H_1(\alpha u)-J_1(\alpha u)\bm
H_0(\alpha u)] \nonumber \\
&&\!\!+[J_0(u)- J_0(\alpha u)]/u\},
\label{hC}
\end{eqnarray}
where $\alpha = a/R$ and $\bm H_n$ is the Struve function.  The 
steady-state temperature increase at the surface is given by
Eq.~(\ref{dTg}), where
\begin{eqnarray}
g_C(\tilde
\rho)&=&\frac{1}{(1-2\alpha/3)}\Big[\ln\Big(\frac{1}{\alpha}\Big)
+\frac{1}{2}\Big(1-\frac{\tilde \rho^2}{3\alpha^2}\Big)\Big],\nonumber \\
&&\;\;\;\;\;\;\;\;\;\;\;\;\;\;\;0\le\tilde \rho\le\alpha,\\
&=&\frac{1}{(1-2\alpha/3)}\Big[\ln\Big(\frac{1}{\tilde \rho}\Big)
+\frac{(\tilde \rho -2\alpha/3)}{\tilde \rho}\Big],\nonumber \\
&&\;\;\;\;\;\;\;\;\;\;\;\;\;\;\;\alpha \le \tilde \rho \le 1,\\
&=&1/\tilde \rho, \;\;\;\;\;\;\;\;\;\;\;\;\tilde \rho
\ge 1.
\end{eqnarray}
At the center of the beam spot,
\begin{equation}
g_{C}(0) = \frac{\ln(1/\alpha)+1/2}{(1\!\!-\!\!2\alpha/3)}.
\label{gC}
\end{equation}

{\bf Model R}.
Another model, proposed by Reimer,\cite{Reimer98} approximates 
the higher density of power dissipation closer to the beam spot by
assuming that the thermal power $P_0$ is deposited uniformly within a
hemisphere of radius equal to $R/2$,
\begin{eqnarray}
P_{v,R}(r)& =& \frac{P_0}{[2\pi (R/2)^3/3]}, r < R/2,\\
&=&0, r> R/2.
\end{eqnarray}
The function $h(u)$ defined in Eq.~(\ref{dTh}) becomes 
\begin{eqnarray}
h_{R}(u)\!\!&=\!\!&
1-(u/2)J_0(u/2)
+J_1(u/2)+(2/u)J_2(u/2)
\nonumber \\
&&\!\!+(\pi u/4)[J_0(u/2)\bm H_1(u/2)-J_1(u/2)\bm H_0(u/2)]. 
\nonumber \\
\end{eqnarray}
The steady-state temperature increase at the surface is given by
Eq.~(\ref{dTg}), where
\begin{eqnarray}
g_R(\tilde
\rho)&=&3-4\tilde \rho^2, \;\;0\le\tilde \rho\le 1/2,\\
&=&1/\tilde \rho,\;\;\;\;\;\;\;\;\;\;\;\;\; \tilde \rho
\ge 1/2.
\end{eqnarray}
At the center of the beam spot,
\begin{equation}
g_{R}(0) = 3.
\end{equation}
Note that the expression for $h_R(u)$ can be obtained from that for
$h_C(u)$ by setting $\alpha = 1$ on the right-hand side of Eq.~(\ref{hC})
and then replacing $u = qR$ by $u/2=qR/2$; similarly, the result for
$g_R(0)$ can be obtained from that for $g_C(0)$ by setting $\alpha = 1$ on
the right-hand side of Eq.~(\ref{gC}) and multiplying by a factor of 2.

{\bf Model B}.
A similar model, proposed by Bresse,\cite{Bresse72} approximates the
density of power dissipation by assuming that the thermal power $P_0$ is
deposited uniformly within a sphere of radius $R/2$ centered a distance
$R/2$ below the surface,
\begin{eqnarray}
P_{v,B}(\rho, z)& =& \frac{P_0}{[4\pi (R/2)^3/3]},\sqrt{\rho^2 +
(z+R/2)^2} < R/2,\nonumber \\
&&  -R < z < 0,\\
 &=&0, {\rm otherwise}.
\end{eqnarray}
The function $h(u)$ defined in Eq.~(\ref{dTh}) becomes 
\begin{eqnarray}
h_{B}(u)= e^{-u/2},
\end{eqnarray}
and the steady-state temperature increase at the surface  is given by
Eq.~(\ref{dTg}), where
\begin{equation}
g_B(\tilde \rho)=\frac{2}{\sqrt{1+4\tilde \rho^2}}.
\end{equation} 
At the center of the beam spot,
\begin{equation}
g_{B}(0) = 2.
\label{gB0}
\end{equation}

\begin{figure}%*****
\includegraphics[width=8cm]{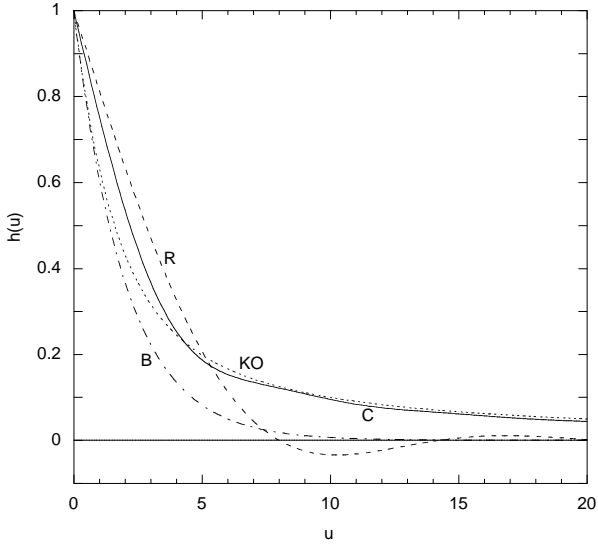}
\caption{$h(u)$, the auxiliary function describing the 2D Fourier transform
$\delta \tilde T(q)$ via Eq.\ (\ref{dTh}), vs
$u = qR$ for the four models discussed in the text: KO (dotted), C(solid),
R(dashed), and B(dot-dashed) when $\alpha =a/R = 0.01$.}
\label{fig1}
\end{figure}

\begin{figure}%*****
\includegraphics[width=8cm]{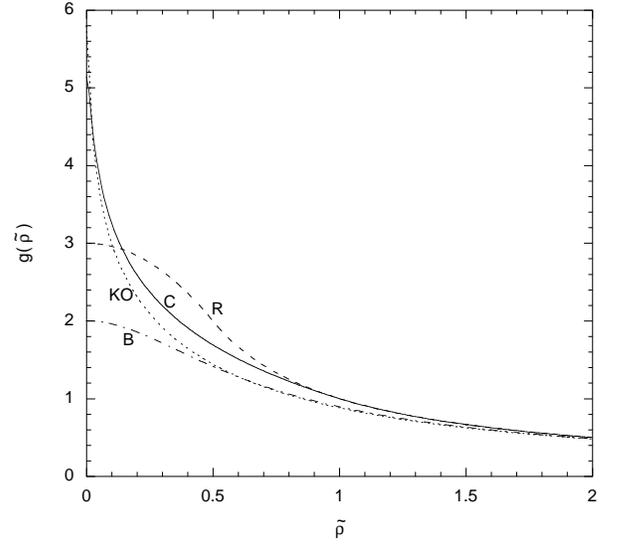}
\caption{$g(\tilde \rho)$,  the auxiliary function describing the
temperature perturbation
$\delta T(\rho)$ via Eq.\ (\ref{dTg}), vs $\tilde \rho = \rho/R$
for the four models discussed in the text: KO (dotted), C(solid),
R(dashed), and B(dot-dashed) when $\alpha =a/R = 0.01$.}
\label{fig2}
\end{figure}

Figures 1 and 2 show plots of the auxiliary functions $h(u)$ vs $u=qR$ and
$g(\tilde \rho)$ vs $\tilde \rho=\rho/R$ for the four models KO, C, R, and
B in the realistic case that the ratio of the e-beam radius $a$ to the
electron range
$R$ is $\alpha = a/R = 0.01.$

For each of these models, the electron-beam-induced shift in the apparent
position of the vortex [see Eqs.\ (\ref{Sf}) and (\ref{dTh})] can be
expressed as 
\begin{equation}
\bm S(\bm \rho_0) = \hat \rho_0\frac{P_0}{\pi \kappa R} \frac{d\Lambda}{dT}
G(\rho_0/R,2\Lambda_0/R),
\label{SG}
\end{equation}
where $G$ is the dimensionless shift,
\begin{equation}
G(\tilde \rho_0,l)= \int_0^\infty\frac{h(u)J_1(\tilde \rho_0 u)}{1+lu}du,
\label{G}
\end{equation}
which is a function of 
$\tilde \rho_0 = \rho_0/R$ and $l = 2\Lambda_0/R = 2\lambda^2(T_0)/Rd$.
For models KO and C, $G$ also depends implicitly upon $\alpha
= a/R$ [see Eqs.~(\ref{hKO}) and (\ref{hC})].

\begin{figure}%*****
\includegraphics[width=8cm]{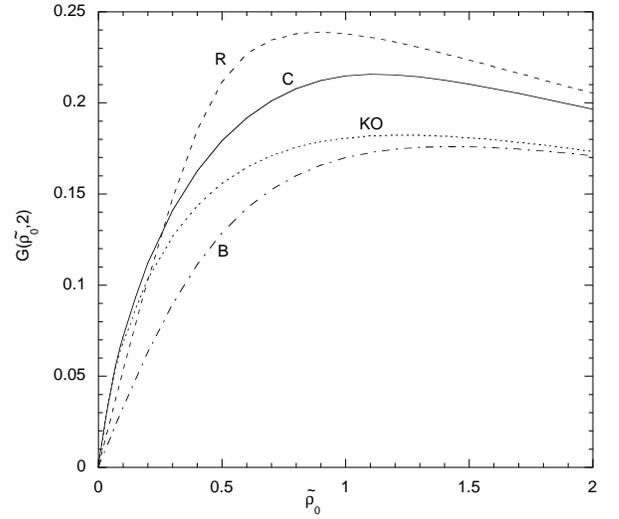}
\caption{Dimensionless shift $G(\tilde \rho_0,2)$ [Eq.~(\ref{G})] vs $\tilde
\rho_0 =
\rho_0/R$ for the four models discussed in the text: KO (dotted),
C(solid), R(dashed), and B(dot-dashed) when $\alpha =a/R = 0.01$ and
$l=2\Lambda_0/R =2$. As seen in Eq.~(\ref{SG}), this function describes the
shift in the apparent position of a vortex pinned at the origin when the
electron beam strikes the film at a distance 
$\rho_0$ from the origin.}
\label{fig3}
\end{figure}

It is straightforward to numerically evaluate the integral in Eq.\
(\ref{G}), and Fig.~3 shows plots of $G(\tilde \rho_0,l)$ vs
$\tilde
\rho_0 =
\rho_0/R$ for
$l=2\Lambda_0/R =2$, which is close to the value calculated for the
experiments of Ref.~\onlinecite{Doenitz06}.  
Although the curves of $G(\tilde \rho_0,l)$ vs
$\tilde
\rho_0$ are qualitatively  similar, the distinct differences in shapes
and maximum slopes may make it possible to
determine which of the four models best fits the experimental
electron-beam-induced shifts in vortex position.
For
$\tilde \rho_0
\ll 1,$ the function
$G(\tilde \rho_0,l)$ is linear in
$\tilde \rho_0$; i.e.,
$G(\tilde \rho_0,l)
\approx \tilde \rho_0 G'(l).$ Since $J_1(x) \approx x/2$ for small $x$,  the
initial slope $G'(l)$ can be seen from 
Eq.~(\ref{G}) to be
\begin{equation}
G'(l)= \frac{1}{2}\int_0^\infty\frac{h(u)u}{1+lu}du.
\label{G'}
\end{equation}

As shown in Fig.~3, the curves for models KO, C,
and R cross near the origin, and for  $\alpha =
a/R = 0.01$ as $\tilde \rho_0 \rightarrow 0,$ the
dotted curve (KO) has the steepest slope $G'_{KO}(2)=1.25,$ followed by the
solid curve (C) with $G'_C(2)=1.07,$  the dashed curve (R)
with $G'_R(2) =0.53$,
and the dot-dashed curve (B) with $G'_B(2)= 0.33$.
As expected from the integrands of Eqs.~(\ref{G}) and (\ref{G'}),
both
$G$ and its initial slope $G'$ decrease monotonically with increasing values
of
$l=2\Lambda_0/R $.  For large
$l$, the initial slope approaches the value $G'(l) = g(0)/2l$, and for the
four models discussed above, this  expression gives,  when $\alpha =
0.01$, 
$G'_{KO}(l)=2.90/l,$  $G'_C(l) =2.57/l$, 
$G'_R(l)=1.5/l,$ and  $G'_B(l)=1/l$.
The largest values of  $G$ and  $G'$ occur for very small values of
$l=2\Lambda_0/R$.  In the limit as $l \rightarrow 0$, the initial slopes are
$G'_{KO}(0) =(1+\alpha -\sqrt{1+\alpha^2})/\alpha^2$ or $G'_{KO}(0) = 99.5$
when $\alpha = 0.01$, $G'_{C}(0) =(6-3\alpha)/(12\alpha-8\alpha^2)$ or
$G'_{C}(0) = 50.1$ when $\alpha = 0.01$, $G'_{R}(0)
=3,$ and $G'_{B}(0)
=2.$

Figure 4 illustrates how the curves of $G(\tilde \rho_0,l)$ vs $\tilde
\rho_0 = \rho_0/R$ depend upon
$l  = 2 \Lambda_0/R$  when model KO is used  
with  beam radius $a$ = 5 nm, screening length $\Lambda_0$ = 0.5
$\mu$m, but electron range
$R$ decreasing (with decreasing beam energy) from $R$ = 100 $\mu$m ($l =
0.01$) to
$R$ = 0.01 $\mu$m ($l = 100$).
Similarly, Fig. 5 (note the different scale for the vertical axis)
shows how the curves of
$G(\tilde
\rho_0,l)$ vs
$\tilde
\rho_0 = \rho_0/R$ depend upon
$l  = 2 \Lambda_0/R$  when model B is used 
with  $\Lambda_0$ = 0.5 $\mu$m and 
$R$ decreasing from $R$ = 100 $\mu$m ($l =
0.01$) to
$R$ = 0.01 $\mu$m ($l = 100$).
The general behavior of  $G(\tilde \rho_0,l)$ vs $\tilde
\rho_0 = \rho_0/R$ is very similar for models KO and B, the chief difference
being that the slope at the origin is much steeper for model KO than for
model B, as discussed above.  Corresponding curves for model C would most
closely resemble those for model KO, and the curves for model R would most
closely resemble those for model B, as can be seen from Fig.~3 

\begin{figure}%*****
\includegraphics[width=8cm]{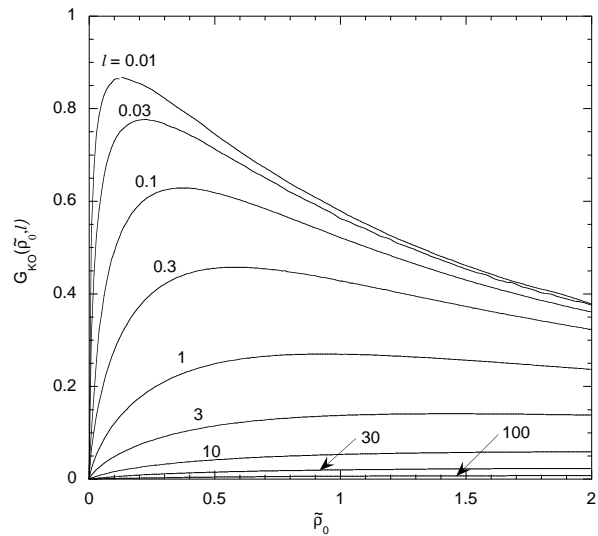}
\caption{Dimensionless shift $G_{KO}(\tilde \rho_0,l)$ vs $\tilde \rho_0 =
\rho_0/R$, calculated using model KO for fixed $a$ and $\Lambda_0$ but
decreasing $R$ as described in the text such that 
$l = 2\Lambda_0/R$ = 0.01, 0.03, 0.1, 0.3, 1, 3, 10, 30, and 100.}
\label{fig4}
\end{figure}

\begin{figure}%*****
\includegraphics[width=8cm]{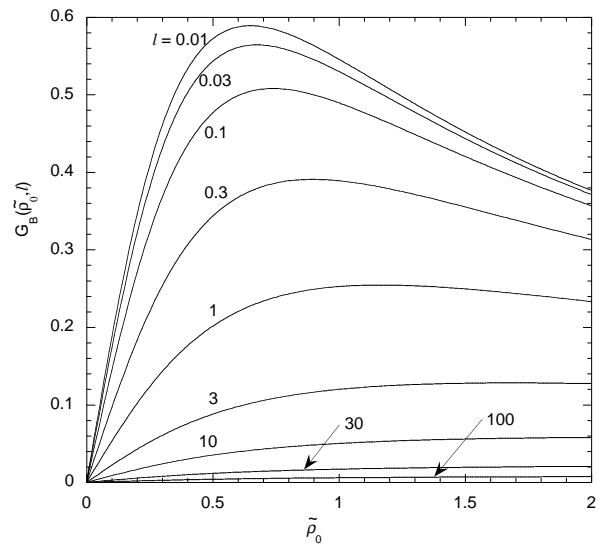}
\caption{Dimensionless shift $G_{B}(\tilde \rho_0,l)$ vs $\tilde \rho_0 =
\rho_0/R$, calculated using model B for fixed $\Lambda_0$ but decreasing
$R$  such that 
$l = 2\Lambda_0/R$ = 0.01, 0.03, 0.1, 0.3, 1, 3, 10, 30, and 100.}
\label{fig5}
\end{figure}

These results indicate that, for a fixed value of $P_0/\pi \kappa R$, the
largest shifts in the apparent vortex position [Eq.~(\ref{SG})] occur for 
values of $l = 2\Lambda_0/R$ of the order of unity or smaller.  The order
of magnitude of the shift when
$\rho_0
\sim R$ is then quickly estimated from Eqs.\ (\ref{K0>}),  (\ref{D}),
(\ref{S}), and (\ref{dT0}) by noting that
$\Lambda_1 \sim (d\Lambda/dT)\Delta T$ over an area of the order of $\pi R^2$
and that $K_0(R) \sim \phi_0/\pi \mu_0 R^2$, which yields a maximum shift
$S(R)
\sim (d\Lambda/dT)\Delta T$, where  $\Delta T
\sim P_0/\pi \kappa R$ is the temperature increment at the center of the
beam spot.  On the other hand, for a fixed electron-beam current $I_b$, the
maximum shift in the apparent vortex position occurs at an intermediate
value of
$l$, 
which can be determined for each of the four models by noting that the
prefactor $P_0/\pi \kappa R$ depends  upon $R =
2\Lambda_0/l$ not only inversely but also implicitly via the incident
electron energy $E_0$, as discussed in Sec.~III.

\section{Discussion}

In this paper I considered a  vortex, pinned at the origin, in a thin film of
thickness $d$ when the local heating produced by a scanning
electron beam focused  at $\bm \rho_0$ locally raises the film's temperature,
decreases the superfluid density, and increases the London penetration
depth
$\lambda(T)$.  Using first-order perturbation theory, I calculated the
resulting vortex-generated supercurrent distribution and the corresponding
distortion of the magnetic-field distribution toward the
e-beam spot.  The resulting expressions for 
the displacement of the center of the field
distribution [Eqs.\ (\ref{SG}) and (\ref{G})] describe how this
shift in apparent position depends upon the thermal power
$P_0$ deposited by the e-beam,
the range
$R$ over which this power is delivered, the beam radius $a$,
the thermal conductivity $\kappa$, the unperturbed
screening length (Pearl length) $\Lambda_0 =\lambda^2(T_0)/d$, the
temperature derivative
$d\Lambda/dT$ of the screening length, and the distance $\rho_0$ between
the e-beam spot and the vortex axis.

I calculated the shift $\bm S(\bm \rho_0)$ using four different models  (KO,
C, R, and B) for the spatial dependence of the thermal power deposited by
the incident electron beam, all models describing the same total thermal
power
$P_0$ deposited within $R$ of the center of the beam spot.
While the results [see Fig.~3] are qualitatively similar for all four
models, the detailed functional form of $\bm S(\bm \rho_0)$ vs $\bm \rho_0$
is strongly model-dependent.  For example, the slope of $\bm S(\bm
\rho_0)$ vs $\bm \rho_0$ at  $\bm \rho_0 = 0$ is much
steeper for models KO and C, which account for the greatly increased
density of power deposition near the center of the beam spot, than for
models R and B,  which assume that the power $P_0$ is distributed
uniformly over much larger volumes.  This suggests that high-resolution
experiments, analyzed with the help of the above theory, could be used to
determine which of the four models gives the best description of the
thermal power density.

To observe e-beam shifts in the apparent position of a pinned vortex,
Doenitz et al.\cite{Doenitz06} used a SQUID with the central hole in the
shape of a long slot.  When a vortex was pinned
in the body of the SQUID, its return flux was measured by the SQUID with
high sensitivity, and any e-beam-induced shift of the vortex's apparent
position toward or away from the slot resulted in a
measurable signal, which was displayed as the intensity on a video display
as the e-beam was rastered across the sample.  Such an image
corresponds to a density plot of
$S_x(x_0,y_0)$ vs $x_0$ and $y_0$. [Note that $S_x(-x_0,y_0) =
-S_x(x_0,y_0).$]   Shown in Fig.~6 is a density plot of $S_x(x_0,y_0)$
calculated using model B for the case $l = 2\Lambda_0/R = 2$, roughly
equivalent to the experimental conditions of Ref.\
\onlinecite{Doenitz06}, where the authors estimated that $R \approx
\Lambda_0
\approx 0.5\; \mu$m.  Figure 6 strongly resembles the vortex images
displayed in  Ref.\
\onlinecite{Doenitz06}.

\begin{figure}%*****
\includegraphics[width=8cm]{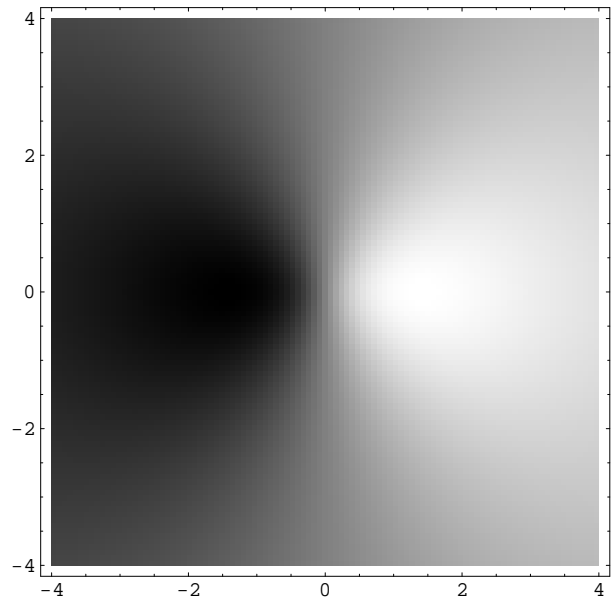}
\caption{Density plot of the $x$ component of the vortex displacement
$S_x(x_0,y_0)$ vs the electron-beam spot position
$(x_0/R,y_0/R)$ on the superconducting film for the case that $l =
2\Lambda_0/R = 2$, calculated using model B described in the text. 
Positive values of
$S_x$ are bright, zero values are grey, and negative values are dark.  The
vortex, at the center of the plot, shifts to the right (bright areas) when
the e-beam strikes to the right of the vortex, and to the left (dark areas)
when the e-beam strikes to the left.  The maximum displacements are shown as
the centers of the bright and dark spots.  A horizontal scan through the
center of the plot [$S_x(x_0/R,0)$ vs $x_0/R$] corresponds to a plot of
$G_{B}(\tilde \rho_0,2)$ vs $\tilde \rho_0$, as in Fig.~3.}
\label{fig6}
\end{figure}

Here is an example of a
numerical evaluation of Eq.~\eqref{SG}.  Assume that the incident
electron energy is  $E_0$ = 10 keV and the beam current is $I_b$ = 7 nA, as
in  Ref.~\onlinecite{Doenitz06}, and that the fraction of the incident
electron energy that is converted into heat\cite{Reimer98} is  $f$ = 0.6,
such that
$P_0$ = 42 $\mu$W.  With a SrTiO$_3$ thermal conductivity $\kappa$ = 18
W/Km  at 77 K (Ref.~\onlinecite{Steigmeier68}) and an electron range $R$ =
0.53 $\mu$m  (Ref.~\onlinecite{Doenitz06}),
$P_0/\pi\kappa R$ = 1.4 K, which is the temperature perturbation at the
center of the beam spot according to model B [see Eqs.~\eqref{dTg} and
\eqref{gB0}].    Assuming
$\lambda(T)=(140 {\rm 
\;nm})/\sqrt{1-(T/T_c)^4}$ and film thickness
$d$ = 80 nm in $\Lambda = \lambda^2(T)/d$, as in
Ref.~\onlinecite{Doenitz06}, one obtains 
$\Lambda_0 =
\lambda^2(T_0)/d
= 0.50\; \mu$m and  $d\Lambda/dT$ = 27.5 nm/K for YBCO ($T_c$ =  91 K)
at $T_0$ = 77 K, such that $l = 2 \Lambda_0/R$ = 1.9.  For model B, the
maximum value of
$G_B(\tilde \rho_0,l)$, which occurs at $\tilde \rho_0 = \rho_0/R$
= 1.41, (see Fig.~5) is 0.181, such that the maximum value of $S_x(x_0,y_0)$
is 7.0 nm  at
$x_0$ = 0.75 $\mu$m and $y_0$ = 0.  This corresponds to the center of
the white spot in Fig.~6.  The center of the black spot in Fig.~6
corresponds to the value $S_x$ = -7.0 nm at $x_0$ = -0.75 $\mu$m and $y_0$ =
0. 

To test which of the above four models
(KO, C, R, or B) is best, similar experiments could be carried out to compare
$S_x(x_0,0)$ vs
$x_0$ for several values of the e-beam radius
$a$  for the same incident electron
energy and accordingly the same range $R$.  
According to models KO and C, 
for the case of an electron
range
$R$ = 0.53 $\mu$m, as in Ref.\
\onlinecite{Doenitz06}, there should
be a significant reduction in the initial slope of
$S_x(x_0,0)$ vs $x_0$ as the beam radius is varied from 
$a$ = 5 nm to 50 nm.  According to models R and B, however, there 
should be no
significant change in slope.

As another test of the above theory, experiments could be carried out over
a wide range of electron energies $E_0$ and corresponding electron ranges
$R$.  Assuming that the range
$R \propto E_0^{1.43}$, as suggested in Ref.~\onlinecite{Reimer98}, Doenitz
et al.\cite{Doenitz06} estimated that $R$ = 0.53 $\mu$m at $E_0$ =
10 keV.  Assuming that this range-energy relation holds for all energies, one
finds the results given in Table I when $\Lambda_0$ = 0.50 $\mu$m.
Comparisons of experimental results with theoretical plots of $S_x(x_0,0)$ vs
$x_0$ obtained from calculations such as those shown in Figs.~4 and 5 would
provide a stringent test of the four models.

\begin{table}
\caption{\label{tab:example}Calculated values of $l=2\Lambda_0/R$,
electron range $R =$ 0.53 $\mu$m$(E_0$/10 keV$)^{1.43}$, and electron-beam
energy
$E_0$ when
$\Lambda_0 = 0.50$ $
\mu$m.}
\begin{ruledtabular}
\begin{tabular}{ccc}
  $l$ & $R$ ($\mu$m) & $E_0$ (keV)\\
0.01 & 100 & 390\\ 
0.03 & 33 & 181\\ 
0.1 & 10 & 78 \\
0.3 & 3.3 & 36 \\
1 & 1 & 16 \\
3 & 0.33 & 7.2 \\ 
10 & 0.1 & 3.1 \\
30 & 0.033 & 1.4 \\
100 & 0.01 & 0.62 \\
\end{tabular}
\end{ruledtabular}
\end{table}

Experiments done at various temperatures above 77 K would provide a
further test of the theory.  In  Ref.~\onlinecite{Doenitz06}, Doenitz et
al.\ estimated the value of $\Lambda_0 =
\lambda^2(T_0)/d
\approx 0.50\; \mu$m for YBCO ($T_c$ =  91 K) of thickness 80 nm at
$T_0$ = 77 K by assuming $\lambda(T)=(140 {\rm  \;nm})/\sqrt{1-(T/T_c)^4}$.
With $R$ = 0.53 $\mu$m, this gives a value of $l = 2\Lambda_0/R = 1.9$.
Increasing the temperature would not only increase the value of $l$ and
decrease the magnitude of $G$ (see Figs.~4 and 5) but also increase
the magnitude of
$d\Lambda/dT$.  For large $l$, the combination of these two competing
effects would lead to an overall increase in the magnitude of
$S_x(x_0,y_0) \propto (1/\Lambda_0)(d\Lambda/dT)$ according to
Eqs.~(\ref{SG}) and (\ref{G}).   With the temperature $T$
approaching
$T_c$, however, it is likely that the vortex would become depinned and follow
the warmer beam spot, such that the above theory, which assumes that the
vortex remains pinned as the electron beam scans over it, would no longer
apply.

Reducing the temperature to lower values would decrease the
value of $l$, thereby increasing the magnitude of $G$. On the other hand,
it is likely that the reduced value of $d\Lambda/dT$ would more than
compensate for this effect and lead to an overall decrease in the magnitude
of
$S_x(x_0,y_0)$.  Going to low temperatures, however,
would put the experiments in a temperature range where the above theory
for the temperature increment involving only the thermal conductivity
$\kappa$  is no longer valid.\cite{Gross94}

It is important to note that the above
theory is expected to be valid for experiments on superconducting films at
liquid-nitrogen temperatures or above. 
At much lower temperatures,  including liquid-helium temperatures, the above
calculations of the temperature perturbation
[Eqs.~(\ref{Frhoz})-(\ref{dT0})] are no longer valid, because it is
then essential to take into account the thermal boundary resistance between
the superconducting film and the substrate.\cite{Clem80,Huebener84,Gross94} 
In such a case, the theory involves an additional length scale, $\eta =
(\kappa d/\alpha_s)^{1/2}$, the thermal healing length, where
$\alpha_s$ is the coefficient of heat transfer from the superconducting film
into the substrate.\cite{Clem80,Huebener84,Gross94}

Although the results given in this paper apply strictly only to a vortex
in an infinite thin film, I expect that when $R \ge \Lambda_0$, Eqs.\
(\ref{SG}) and (\ref{G}) also apply to good approximation to a vortex
in a thin film with {\it finite} lateral dimensions, provided that the
distance of the vortex from the edge of the film is more than a few
$R$.  This is because the current-density perturbation is largest within
an area of order $R^2$ when the electron beam is roughly a distance $R$
from the vortex.  However, when the distance of the vortex from the edge
of the film is approximately $R$ or smaller, the above
calculation would have to be redone, taking into account how the
supercurrent is modified near the film edge.  

\acknowledgments

I thank D. Koelle for stimulating discussions and many helpful suggestions
during the course of this research.  This work was
supported by Iowa State University of
Science and Technology under Contract No.\ W-7405-ENG-82.


\begin{thebibliography}{99}
\bibitem{footnote} I assume here that
$\delta T(\bm \rho)$ depends only upon $\rho = |\bm \rho|$, such that
$\delta {\tilde T}(\bm q)$ depends only upon $q = |\bm q|$.
\end{thebibliography}
\end{document}